\pdfoutput=1

\documentclass[11pt]{article}
\usepackage{xcolor}

\usepackage[final]{acl}

\usepackage{times}
\usepackage{latexsym}
\usepackage{mathtools}
\usepackage{tikz}
\usepackage{multirow}
\usepackage{colortbl}
\usepackage{booktabs}
\usepackage{multirow}

\usepackage[T1]{fontenc}

\usepackage[utf8]{inputenc}

\usepackage{microtype}

\usepackage{inconsolata}

\usepackage{graphicx}

\title{Dia-Lingle: A Gamified Interface for Dialectal Data Collection}

\newcommand*\samethanks[1][\value{footnote}]{\footnotemark[#1]}
\author{
 \textbf{Jiugeng Sun\textsuperscript{1}},
 \textbf{Rita Sevastjanova\textsuperscript{1}\thanks{Equal contribution.}},
 \textbf{Sina Ahmadi\textsuperscript{2}\samethanks[1]},
 \\
 \textbf{Rico Sennrich\textsuperscript{2}},
 \textbf{Mennatallah El-Assady\textsuperscript{1}}
\\
 \textsuperscript{1}Department of Computer Science, ETH Zürich, Switzerland 
 \\
 \textsuperscript{2}Department of Computational Linguistics, University of Zürich, Switzerland
\\
 \small{
   \textbf{Correspondence:} \href{mailto:jiusun@ethz.ch}{jiusun@ethz.ch}
 }
}

\begin{document}
\definecolor{mygreen}{HTML}{d9ead3}

\newcommand{\circnum}[1]{%
    \tikz[baseline=(char.base)]{
        \node[shape=circle, draw, fill=mygreen, inner sep=0.93pt] (char) {#1};
    }%
}
\maketitle
\begin{abstract}
Dialects suffer from the scarcity of computational textual resources as they exist predominantly in spoken rather than written form and exhibit remarkable geographical diversity. Collecting dialect data and subsequently integrating it into current language technologies present significant obstacles. Gamification has been proven to facilitate remote data collection processes with great ease and on a substantially wider scale. This paper introduces Dia-Lingle, a gamified interface aimed to improve and facilitate dialectal data collection tasks such as corpus expansion and dialect labelling. The platform features two key components: the first challenges users to rewrite sentences in their dialects, identifies them through a classifier and solicits feedback, and the other one asks users to match sentences to their geographical locations. Dia-Lingle combines active learning with gamified difficulty levels, strategically encouraging prolonged user engagement while efficiently enriching the dialect corpus. Usability evaluation shows that our interface demonstrates high levels of user satisfaction. We provide the link to Dia-Lingle: \url{https://dia-lingle.ivia.ch/}, and demo video: \url{https://youtu.be/0QyJsB8ym64}. 
\end{abstract}

\section{Introduction}

\begin{figure*}[t]
  \centering
  \includegraphics[width=\linewidth]{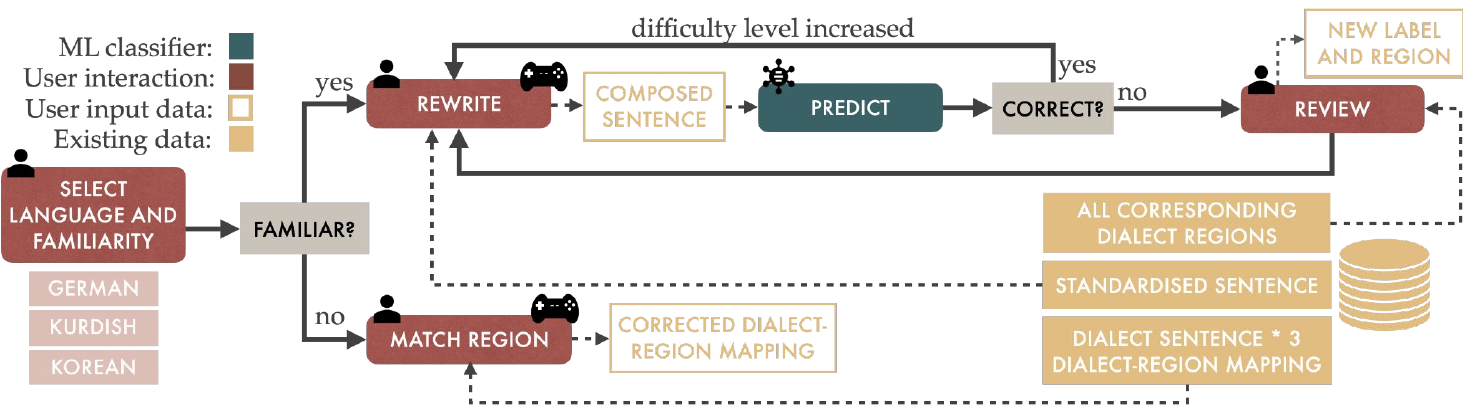} 
  \caption {Illustrative workflow of Dia-Lingle with colour-encoded components for clarity.}%
    \label{fig:pipeline}
\end{figure*}

Dialects present unique challenges for computational linguistics due to their scarcity of textual resources, with limited datasets and digital documentation tools dedicated to their study and analysis~\cite{DBLP:journals/corr/abs-2401-05632}. This digital resource gap, which affects low-resourced languages broadly, reflects systematic issues like cultural marginalisation. Consequently, language communities are often unable to fully benefit from advancements in language technology, raising concerns about the potential erasure of linguistic and cultural diversity in the AI models \citep{grutzner-zahn-etal-2024-surveyingg}.

In the current AI landscape, where data plays a central role---particularly in the advancement of large language models (LLMs)---ensuring fair representation of dialects remains a pressing challenge. Paradoxically, the pragmatic implementation of language standards often forces dialects into standardised written forms for efficiency, which fundamentally undermines authentic dialectal representation and complicates data collection efforts~\cite{auer2005europe}. Previous approaches in natural language processing (NLP) have attempted to address this challenge by leveraging syntactic atlases, structured questionnaires, and direct annotation by native speakers, though these annotations remain limited in scope, as discussed by \citet{alam-etal-2024-codett}. 

In this vein, we design and develop Dia-Lingle, an interactive dashboard that aims to collect dialectal data through two gamified components, as showcased in \autoref{fig:pipeline}. In the first component, dubbed `\textbf{Quiz}', users rewrite standardised sentences, i.e. sentences in the standard variety of the language,
in their dialects. A dialect identification classifier predicts the geographical origin of these rewritten sentences followed by a visualisation of these dialect regions to users, soliciting their feedback both to refine classification accuracy and to build a more comprehensive dialectal dataset. In addition, we integrate active learning (AL) techniques for the strategic recommendation of sentences, selecting suitable ones based on model uncertainty measurements and explicit user feedback. This AL approach is embedded within a difficulty level setting, where challenges escalate based on user proficiency and model performance, thereby enhancing engagement and extending participation in the data collection cycle. The second component, dubbed `\textbf{Match}', requires users to match example sentences to the geographical areas where they are spoken. 
A comprehensive usability study is conducted, revealing user satisfaction from interface design to overall concept. 

To summarise, our main contributions are: (1) introducing gamification for dialectal data collection, while (2) integrating AL for sentence recommendation, and (3) proposing a visualisation and interaction approach for dialect region coverage.

\section{Related Work}

\paragraph{Dialects in NLP} NLP tasks such as machine translation %
systems typically require training datasets comprising tens or even hundreds of millions of sentences. Datasets of this magnitude are available only for a small number of highly resourced languages \cite{haddow-etal-2022-survey}. Despite the increasing attention to addressing the computational processing of low-resourced language varieties and dialects~\cite{Zampieri_Nakov_Scherrer_2020, nekoto-etal-2020-participatory}, efficiently collecting dialectal data without substantial time and financial investment remains a hurdle~\cite{magueresse2020lowresourcelanguagesreviewpast,deutsch2025wmt24}. Previous studies rely on a range of approaches, from extracting data from syntactic atlases~\cite{DBLP:conf/konvens/ScherrerR10}, movie dialogues~\cite{DBLP:conf/coling/AhmadiJAA24} and social media posts~\cite{DBLP:conf/emnlp/CamachoColladosRRULABALC22} to more costly manual annotation~\cite{DBLP:journals/corr/abs-2102-11000}.

\paragraph{Gamification in NLP} Definitions of gamification vary considerably, typically emphasising either game elements and mechanics or the process of gaming and playful experiences in serious contexts \cite{gamification_Deterding, zichermann2011gamification, SEABORN201514, KRATH2021106963}. In this paper, we primarily adopt the conceptualisation proposed by \citet{gamification_Deterding}, which defines gamification as the use of game elements in non-game contexts. Studies have demonstrated that gamification can facilitate remote data collection with greater ease and on a substantially wider scale, yielding ecologically valid and robust data~\cite{godard-etal-2018-loww}. Yet, this methodology remains underutilised in applied linguistics \cite{10.1093/applin/amad039}, with limited research primarily focused on such as second-language acquisition \cite{app12031643}, or dialogue data collection \cite{pairmeup, asher-etal-2016-discourse, ogawa-etal-2020-gamification} and text-labelling \cite{10.1145/2448116.2448119}. 

\paragraph{Data Collection using AL} AL is a specialised form of semi-supervised machine learning that incorporates users into the loop, querying them for label information to enhance classifier training performance \cite{Olsson2009ALS}. The core component of AL is the candidate selection strategy, which aims to identify instances that would contribute most significantly to the model's learning progress \cite{8019851}. Previous work has demonstrated the application of AL in language identification (LID); for example, \citet{lippincott-van-durme-2021-active} establish that utilising negative evidence can improve the performance of simple neural LID models. 

To fill the current gaps in effective language-agnostic dialectal data collection, we develop a gamified interface that involves players as participants who help improve and enrich current dialectal datasets thanks to their feedback. Additionally, we integrate AL with gamified elements to create an improved candidate selection strategy specifically for data aggregation scenarios.%

\section{Methodology}

Our methodology integrates gamification by leveraging player feedback as its core data collection mechanism. The system strategically selects sentences based on model uncertainty, refines dialect identification through user input, and enhances engagement through geographical visualisation. 

\subsection{Users}

Dia-Lingle is designed for three distinct user groups: (a) dialect community members with basic to proficient understanding of specific dialects or dialect families; (b) general linguists interested in analysing the classifier performance and providing sophisticated feedback; and (c) learners in educational contexts to engage with Match component to better understand dialectal variations. This multi-audience approach ensures the platform serves both data collection and model refinement purposes. %

\subsection{Data Structure}
Dia-Lingle utilises two different types of datasets that serve distinct functions within the system. The first dataset comprises archival general information about each dialect, which enriches the user experience by providing contextual background and enhances the visualisation component of the interface. The second dataset contains a corpus of sentences written in various dialects, which provides the training data for the dialect identification classifier and forms the foundation of an expanding resource that will be aggregated for future downstream machine learning tasks like dialect-specific language modelling. Both datasets are continually refined through user feedback embedded in the interface, creating a dynamic and increasingly comprehensive resource for dialectal research.

\subsubsection{Dialect Representation}
\label{sec:Dialect_representation}

\begin{figure}[t]
  \includegraphics[width=\columnwidth]{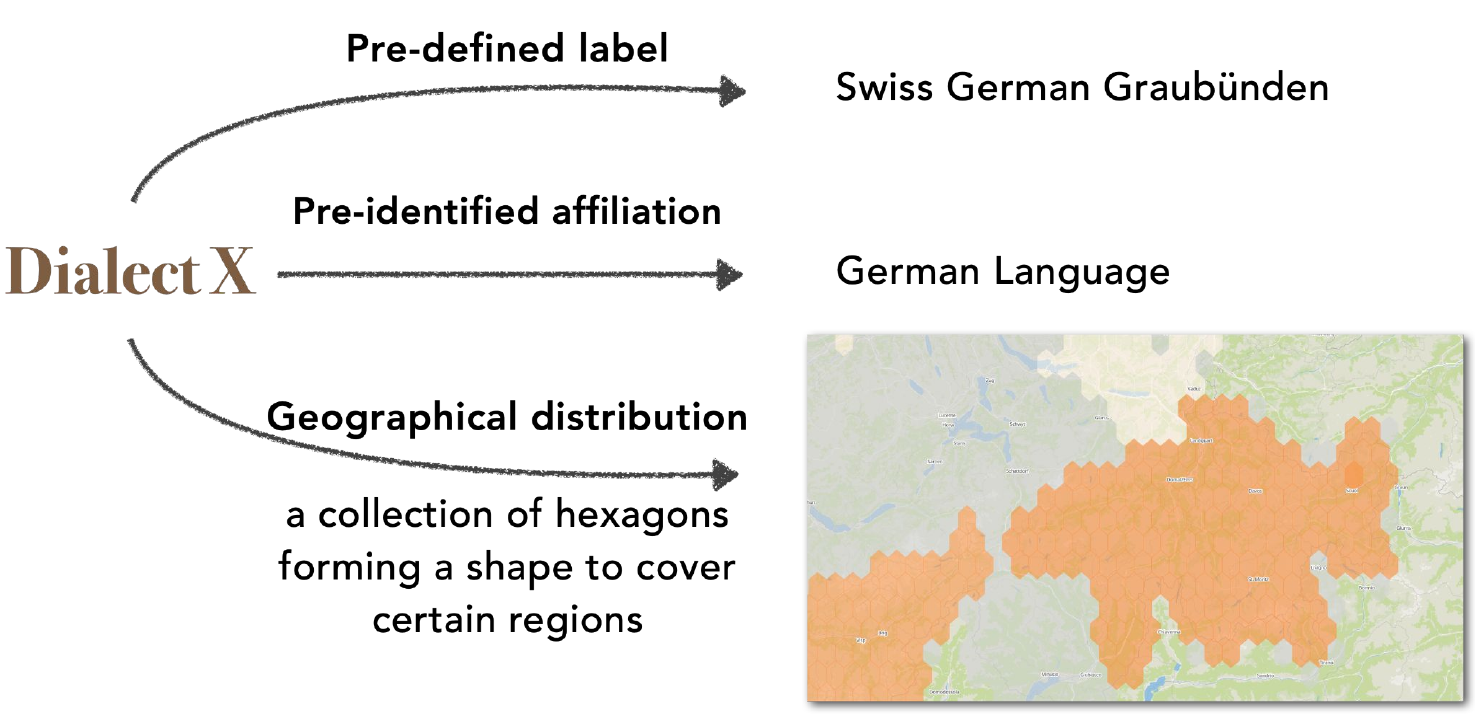}
  \caption{Illustration of dialect representation in Dia-Lingle
  using Dialect X as an example spoken primarily in the Graubünden 
  region of Switzerland. }
  \label{fig:dialect_representation}
\end{figure}

Our approach represents dialects from data science and geographical perspectives rather than discussing their political factors or sociolinguistic impact. Each dialect in our database is assigned three attributes: a pre-defined label, a pre-identified affiliation (i.e., to which macro-language family it belongs), and a geographical distribution. The geographical distribution is represented as a polygon which is formed by a collection of hexagons that covers the regions where the dialect is predominantly spoken. The hexagon sizes are tailored to correspond with the geographical coverage of each language on the world map, with smaller hexagons representing languages spoken in more limited regions, so that the collective hexagon arrangement forms a polygon shape reflecting each language's dialectal distribution. 
\autoref{fig:dialect_representation} illustrates this approach with an example of \textit{Dialect X}, which is primarily spoken in the Graubünden region of Switzerland.

\subsubsection{Corpus}

\begin{figure}[t]
  \includegraphics[width=\columnwidth]{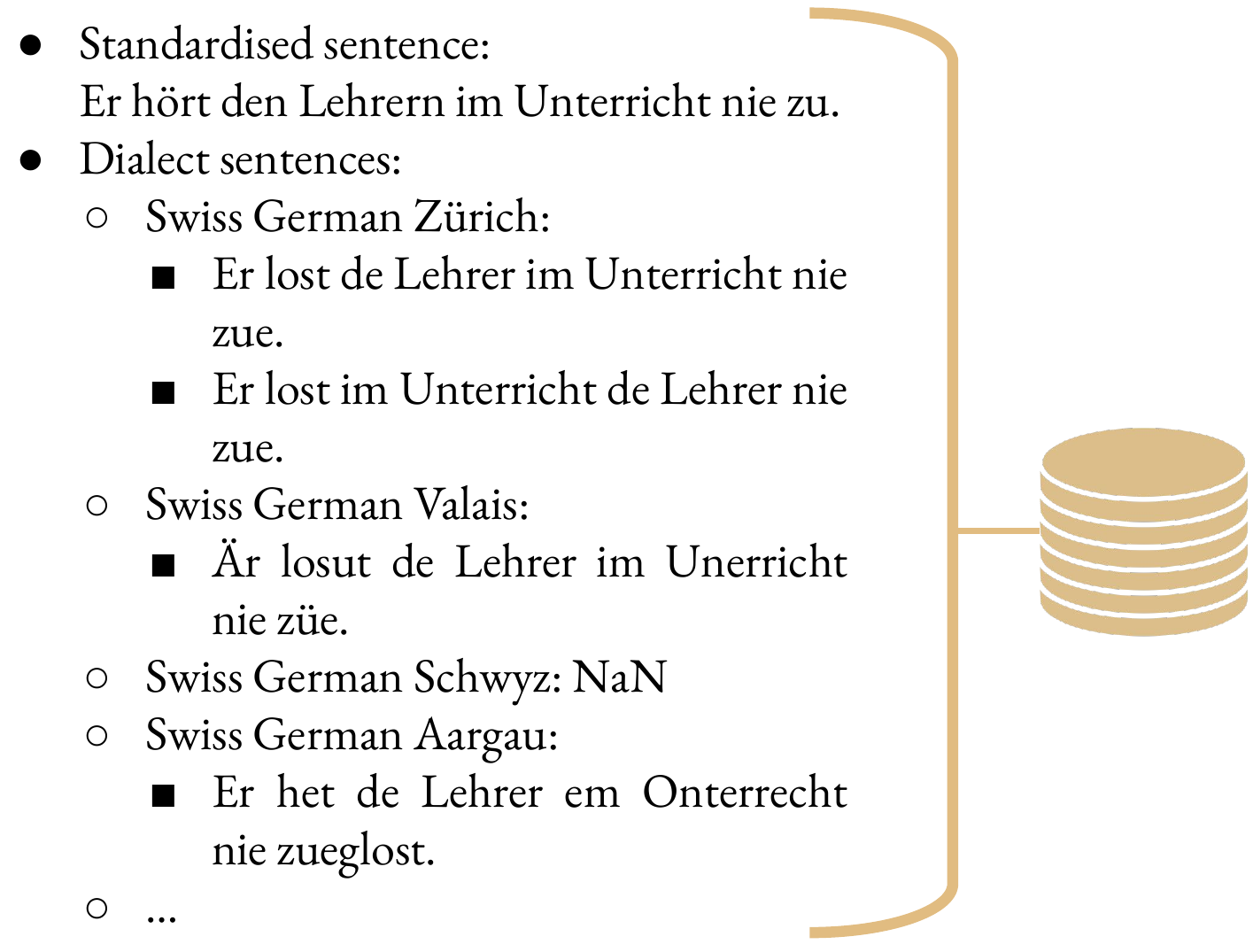}
  \caption{Illustration of a parallel sentence group in Swiss German. There is one standardised sentence and multiple dialect sentences that convey the exact same meaning. }
  \label{fig:language_group}
\end{figure}

The corpus comprises a collection of parallel sentences, with each data entry containing a standardised language version alongside dialectal variations that convey the same meaning. All sentences are labelled using the same pre-defined labels established in the dialect representation dataset as discussed in Section~\ref{sec:Dialect_representation}. Importantly, a dialectal sentence may belong to more than one dialect category, reflecting the natural overlap in linguistic features across related dialect groups. We present an example on one of the possible dialect sentence groups in Swiss German to better illustrate the corpus structure in \autoref{fig:language_group}. 

Currently, our corpus encompasses five language families: Swiss German (2,760 parallel sentence groups with eight dialect variations), Kurdish (300 parallel sentence groups with four dialect variations), Korean (51,963 parallel sentence groups with one dialect variation), Japanese (500 parallel sentence groups with twenty-one dialect variations) and Romansh (113 parallel sentence groups with five dialect variations). These datasets were established from existing resources including SwissDial \cite{swissdial}, CODET \cite{alam-etal-2024-codett}, Jejueo Dataset \cite{park-etal-2020-jejueoo}, and CPJD corpus \cite{takamichi-saruwatari-2018-cpjd}. For Romansh, we relied on a few stories from Storyweaver\footnote{https://storyweaver.org.in} and their translations by the \textit{Romanische Kindergeschichten} application (all licensed CC BY 4.0).

\subsection{Classification}
\label{sec:classification_model}
For dialect classification, Dia-Lingle relies on Meta's \texttt{fasttext}~\cite{mikolov2018advances} which provides character-level word embeddings, sentence classification and most importantly, language identification for 157 languages. For the task of dialect classification, we train models for the selected varieties in our corpus. During training, we use automatic hyperparameter optimisation, constraining the model size to a maximum of 2 MB and limiting the autotune duration to 500 seconds. The model performance was evaluated using an 80/20 train-test split. The $F_1$ scores for different dialects are as follows: 0.878 for Swiss German, 0.546 for Kurdish, 0.973 for Korean, 0.446 for Japanese and 0.842 for Romansh. Some language families still have relatively low $F_1$ scores due to limited corpus resources. Given that the interface focuses on integrating human feedback for data collection, model performance is less prioritised at this moment.

\begin{figure*}[t]
  \includegraphics[width=\linewidth]{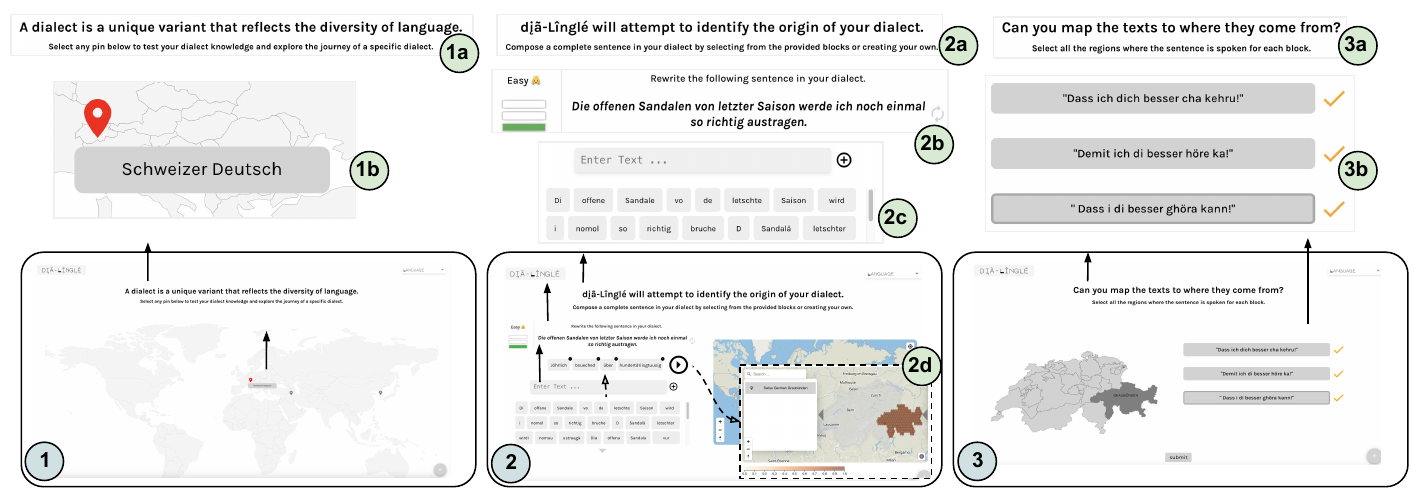} 
  \caption{Simplified overview of Dia-Lingle interface design, detailed in Section~\ref{sec:interface_design}. Major components are enlarged for visibility and labelled with circled numbers for reference. \autoref{sec:appendix} provides additional images of other stages.}
  \label{fig:interface_design}
\end{figure*}

\subsection{Gamification on Progressive Levels}
\label{sec:AL_on_model}
Game interface design encompasses a diverse array of design patterns, among which the Levels are a prominent exemplar \cite{gamification_Deterding}. Levels are a mechanism to provide users with progression within the system \cite{10.1145/3611048}. The gamified strategies include offering rewards for completing levels \cite{8820847} or increasing difficulty \cite{SEABORN201514}. \citet{https://doi.org/10.1111/medu.12190} develop a voluntary online quiz system and reveal through post-questionnaires that 96\% of participants attribute their participation to difficulty level settings. In Dia-Lingle, we implement a three-tiered difficulty level setting: Easy, Normal, and Hard. Users initially receive a standardised sentence from the Easy category when entering the rewriting page. Either they can modify the difficulty level themselves, or upon expressing satisfaction with the classification result and dialect visualisation, Dia-Lingle increases the difficulty level to encourage further participation.

\subsection{Uncertainty-based Sample Selection}
During data aggregation process, selecting sentences for users to rewrite requires careful consideration. Our objectives are twofold: to collect valuable and preferably unseen dialectal data, whilst encouraging user engagement. 
In each interaction round, we define a difficulty score $D(s)$ for a standardised sentence $s$ as:
\vspace{-3pt}
\begin{equation}
  \label{eq:difficulty_score}
  D(s) = \sum_{k\in K}H(k)
\end{equation}
\vspace{-5pt}
\\
where $k$ denotes the $k$-th dialect variation class and $H(k)$ represents the classification entropy of the $k$-th class for a standardised sentence $s$. We define the parallel dialect sentence group of sentence $s$ to be $G_s$, where $G_{s,k}\subset G_s$ denotes the subset of dialect sentences with label $k$. Furthermore, we define $K$ as the set of all possible dialect variations and $C\subseteq K$ as the set of labels that actually appear in $G_s$. We define $H(k)$ to be:
\vspace{-3pt}
\begin{equation}
\small
  \label{eq:entropy}
  H(k) = \begin{cases}
        -\frac{1}{|G_{s,k}|}\sum_{n\in G_{s,k}}\sum_{i\in K}p_{n_i}\mathrm{log}(p_{n_i}) & k\in C \\
        H_{\mathrm{max}}(k) = - \sum_{i\in K} \frac{1}{|K|}\mathrm{log}(\frac{1}{|K|}) & k\notin C \\
    \end{cases}
\end{equation}
\vspace{-5pt}
\\
where $p_{n_i}$ is the probability of the dialect sentence $n$ belonging to the $i$-th class. This score increases when certain dialect variations are missing from the parallel group or when the model struggles to confidently classify the dialect sentences. Standardised sentences are categorised by difficulty score: Hard (top 20\%), Normal (middle 60\%), and Easy (bottom 20\%).
As users interact with the Dia-Lingle interface, we continuously expand the corpus by adding more dialect sentences to the corresponding group $G_s$, potentially expanding the label set $K$ and updating the dialect classifiers. %

\section{Interface Design}
\label{sec:interface_design}
Dia-Lingle's interface, depicted in \autoref{fig:interface_design}, guides users through several stages: Entry, Choice, Quiz, Review and Match. These stages encompass two distinct and separate gamified components that users can engage with. In our commitment to engage with local dialect communities, we have prioritised multilingual support across the interface. %
Beyond the interactions occurring within the main user flow, the interface incorporates numerous additional interactive components to enhance usability, which include pop-up windows to prevent unintended actions or onboarding guidelines.

\subsection{Entry}
Upon entering Dia-Lingle, users are greeted in a welcome page featuring a centred title as shown in \autoref{fig:interface_design} \circnum{1a}. 
Beneath is the navigation area, a world map displays several pins representing the current language families supported by the system. Users progress by selecting any of these pins as depicted in \autoref{fig:interface_design} \circnum{1b}.

\subsection{Choice}
\label{sec:choice}
\begin{figure}[b]
  \includegraphics[width=\columnwidth]{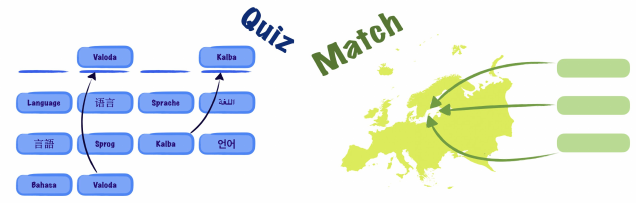}
  \caption{Illustration of the two gamified components (Quiz and Match) as they appear on the Choice page.%
  }
  \label{fig:choice}
\end{figure}

After clicking any pin on the world map, users enter the Choice page divided by a vertical dashed line. The two blocks positioned on either side represent distinct paths leading to two different gamified components as shown in \autoref{fig:choice}. This page requires users to specify whether they are familiar with the language they have selected. Their response determines which of the two major gamified components they will access.

\subsection{Quiz}
\label{sec:quiz}
Users proficient in the language access a page to rewrite sentences in their dialect. The interface includes a title (\autoref{fig:interface_design} \circnum{2a}), a rewriting dashboard with the standardised reference sentence and difficulty toggle (\autoref{fig:interface_design} \circnum{2b}), and below these, a text input field with word suggestions from the dialect corpus (\autoref{fig:interface_design} \circnum{2c}).
Additionally, users can compose their dialectal sentence either by typing directly or by selecting suggested words. The sentence is assembled using interactive blocks that allow users to rearrange word order or remove elements as needed. The interface accommodates left-to-right and right-to-left writing systems.
Once the rewriting is complete, users submit their sentence, triggering the classifier to identify the dialectal labels of the input. The classification results are then visualised on a map as shown in \autoref{fig:interface_design} \circnum{2d}.

\subsection{Review}
\label{sec:review}
The moment the map visualises the classification result, the rewriting interface on the left disappears, replaced by a question asking users whether they believe the identification is accurate. If users are satisfied with the prediction, the interface increases the difficulty level and redirects them back to the rewriting interface described in Section~\ref{sec:quiz}. Otherwise, the interface initiates a feedback collection process. It firstly displays a comprehensive list of all currently archived dialects that belong to the selected language. Users may then select the option that best matches their dialect, or choose \textit{none of the above} to create their own entry. Importantly, regardless of which option users select, they can only modify the geographical regions associated with the dialect. They cannot alter either the dialect labels or the language affiliations detailed in Section~\ref{sec:Dialect_representation}. This restriction acknowledges the sensitivity surrounding dialect and language naming conventions. 

Following the users' selection, the interface activates a lasso tool enabling them to colour the areas on the map where they believe the dialect is spoken. Users can add or remove highlighted small hexagons to form a new polygon representation through either group selection or individual clicking. Upon submission of these geographical modifications, users are redirected back to the rewriting page without an increase in difficulty level. 

The detailed illustrations of the Review session are included in \autoref{sec:appendix}.

\subsection{Match}
Users who express unfamiliarity with the language in Section~\ref{sec:choice} are directed to an alternative component. In this gamified component, the interface presents three sentences sequentially as illustrated in \autoref{fig:interface_design} \circnum{3b}, asking users to highlight regions where they believe each sentence might be spoken. Once the reference answers are revealed, users have the chance to correct these mappings if they disagree with the suggested geographical distributions. 

To maintain simplicity and intuitive gameplay, this matching exercise operates at the administrative division level (such as cantons or provinces) rather than using the more granular hexagonal mapping system. This design choice ensures Match component remains accessible to people of all backgrounds, regardless of their linguistic expertise. 

\section{Evaluation}
\label{sec:evaluation}

We conducted a user study on the interface design discussed in Section~\ref{sec:interface_design}. 
The comprehensive experimentation on the interface was carried out through surveys containing basic demographic data questions and questions following the System Usability Scale (SUS) format \cite{sus}. We also conducted usability studies to gain a better understanding of the participants' comprehension of the entire workflow and any potential obstacles they might encounter whilst interacting with Dia-Lingle. The contents of the survey and the detailed process of the usability test are illustrated in \autoref{sec:appendix}. 

\begin{figure}[t]
  \centering
  \includegraphics[width=\columnwidth]{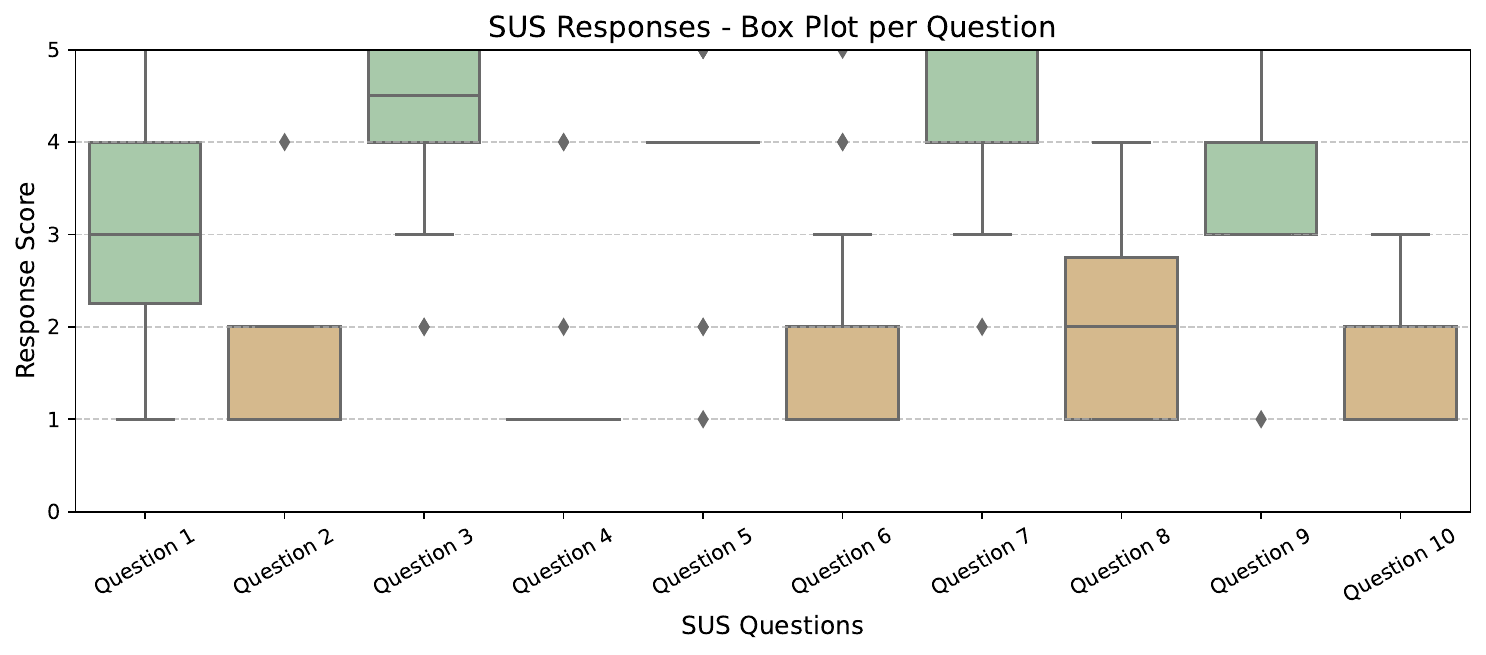}
  \includegraphics[width=\columnwidth]{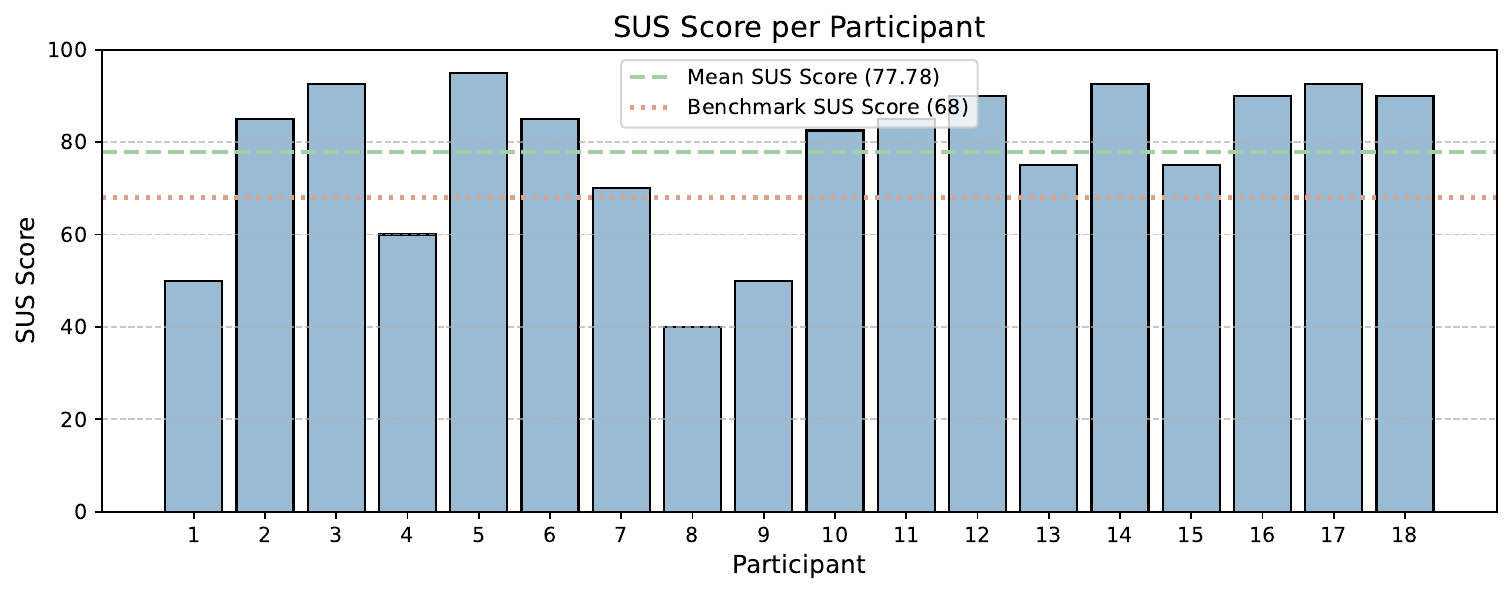}
  \caption{Analysis of SUS results for Dia-Lingle. Top: Box-plot of SUS scores by individual questions, with green boxes indicating higher values are better and brown boxes indicating lower values are better. Bottom: Bar chart showing individual user SUS scores. SUS questions are provided in \autoref{sec:appendix}, \autoref{tab:survey}.}
  \label{fig:sus_main}
\end{figure}

\paragraph{Quantitative Evaluation} In total, 18 participants interacted with the interface and completed the survey. Among the participants, 13 people reported using their native languages to explore the interface. We calculate the overall average SUS score to be 77.78, as defined by \citet{sus}. Detailed results of SUS scores by question and individual user can be found in \autoref{fig:sus_main}. %

In the SUS survey, most questions received highly satisfactory results. However, the first question \textit{I think I would use this interface frequently} produced mixed results. 
Although most participants expressed appreciation for the difficulty levels that motivated them play for more rounds, some preferred elements like point collection or competition for a long-lasting engagement. 
This supports previous research indicating that game element preferences are subjective \cite{10.1145/2967934.2968082}, highlighting the need for more personalised solutions.
Based on the demographic data gathered, non-native speakers highly recognised the educational value. Among native speakers, Swiss German speakers reported higher satisfaction compared to Korean participants, attributable to the more comprehensive pre-established Swiss German dataset, which fostered greater trust and gave users more confidence in utilising the interface. 

\paragraph{Qualitative Evaluation} We conducted usability studies with four participants in total: three Swiss German native speakers and one Korean native speaker. The study instructions and a detailed analysis of identified strengths and problems during the evaluations are provided in \autoref{sec:appendix}. Overall participants exhibited a notably positive reception to the interface's core concept, with particular emphasis on the gamification that substantially increased the platform's attractiveness and user engagement to them. 

One key finding is the critical importance of correctly classifying the first rewritten sentence. Misclassification at this stage raised doubts among users, significantly reducing their confidence in proceeding. Another major finding during the studies is that while the difficulty scores described in Section~\ref{sec:AL_on_model} are based on the ML model’s uncertainty and missing data entries, participants reported that higher difficulty levels actually corresponded to increased complexity in rewriting standardised sentences to them. This was due to either the sentences themselves being inherently more challenging or the suggestion text blocks becoming fewer. Although it remains unclear whether the backend AL algorithm or the frontend difficulty level visualisation played a more significant role in user perception, the overall gamification approach successfully increased participant engagement. These findings offer valuable insights for the future development of Dia-Lingle.

\section{Conclusion}
\label{sec:conclusion}

Dia-Lingle introduces a novel gamified approach to dialectal data collection, addressing the critical challenge of resource scarcity for dialects in computational linguistics. By integrating active learning with gamification elements, our platform creates an engaging environment that encourages prolonged user participation while systematically enriching dialectal corpora. The two-component design--Quiz for dialect rewriting and identification, and Match for geographical mapping—provides complementary pathways for data collection tailored to different user knowledge levels. Our usability evaluation demonstrates high levels of user satisfaction with the interface design and overall concept. The integration of difficulty progression, geographical visualisation, and interactive feedback mechanisms has proven effective in sustaining user engagement. Key insights from our evaluation highlight the importance of accurate initial classification in establishing user trust and the value of strategically increasing challenge levels to maintain participation. 

Dia-Lingle represents a significant step toward more inclusive language technology by creating pathways for communities to contribute directly to the resources that will power future NLP applications. This, consequently, also contributes to the broader goal of preserving linguistic diversity.

\section*{Limitations and Future Work}
\label{sec:future_work}

Currently we only support five different languages on the dashboard. In future research, we aim to broaden the scope of language coverage in dialect identification. As migration and language contact influence speakers, sentences may exhibit characteristics of multiple dialects within a single language or even features from multiple languages. Beyond probability distributions, more sophisticated visualisation methods are needed to effectively represent dialectal mixtures. Additionally, developing machine learning models capable of encoding and detecting such patterns remains an open research challenge. Furthermore, dialects are an integral part of language diversity, and they also exist in spoken rather than written form. Incorporating audio input alongside textual data is crucial, as certain dialectal variations manifest at the phonotactic level. Last but not least, currently evaluation and promotion work on the interface is biased on the fact that most of the participants are highly-educated people and they at least hold a bachelor degree. In the future we are aiming for conducting more formal experiments on the local communities and adapt the interface to satisfy their needs and concerns.

\section*{Ethical Considerations}

In developing the Dia-Lingle platform, we have carefully considered ethical implications of dialectal data collection. Our interface content has been designed to exclude sensitive, discriminatory, or potentially offensive language. All text elements appearing throughout the interface have been reviewed and verified by native speakers to ensure cultural appropriateness and linguistic accuracy for most of the language options. Currently, we offer ten language options: English (the reference language), French, Italian and Spanish (machine-translated), and German, Simplified Chinese, Japanese, Korean, Latvian and Kurdish (machine-translated and subsequently verified by native speakers).

To prevent potential conflicts related to dialectal identity, the platform does not permit users to arbitrarily modify dialect names, which helps avoid contentious naming disputes that could arise from different sociolinguistic perspectives.

\section*{Acknowledgments}
\looseness=-1
We want to express our genuine gratitude to Eunjung Cho, Yuxuan Zhang, Raphaël Baur and Michitatsu Sato for externally helping us verify Korean, German and Japanese machine-translated texts in Dia-Lingle. In particular, Eunjung Cho and Michitatsu Sato have also helped us create the visualisation on Korean and Japanese dialects respectively. This work was supported by the Swiss National Science Foundation through the \mbox{MUTAMUR} project (no. 213976) and the Personalized Visual Analytics project (no. 10003068).

\bibliography{acl_latex}

\clearpage
 \onecolumn
\appendix

\section{Appendix}
\label{sec:appendix}

\begin{figure*}[!h]
  \includegraphics[width=\linewidth]{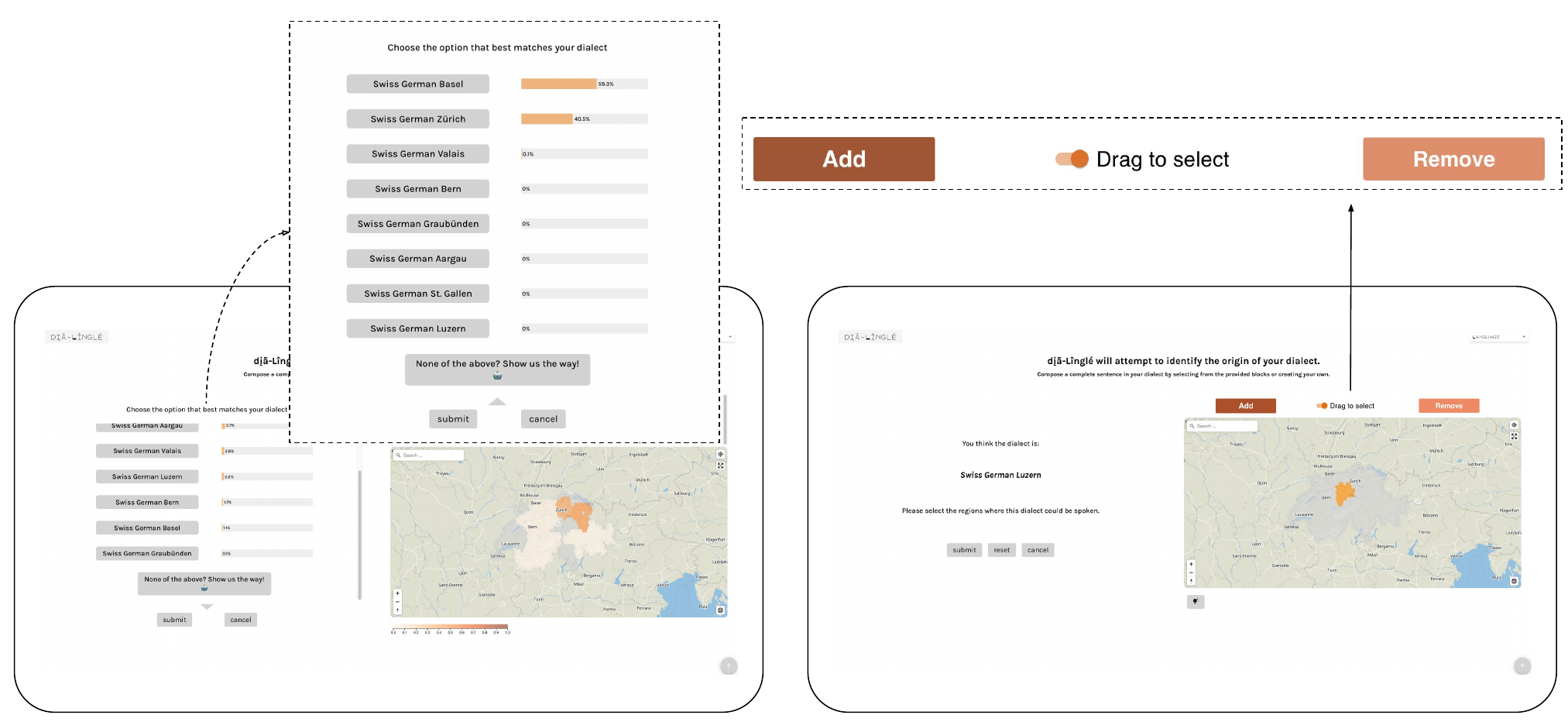} 
  \caption{Overview of the Dia-Lingle interface design at the Review stage, detailed in Section~\ref{sec:review}. The interface displays a list of all possible dialects and offers a lasso tool enabling users to define more accurate dialect regions.}
  \label{fig:review}
\end{figure*}

In this section, we include the detailed overview of the Review stage from the Dia-Lingle interface as shown in \autoref{fig:review}. We also include the content of the survey questionnaire in \autoref{tab:survey} and instructions for the usability studies in \autoref{tab:u_s}. Usability studies were conducted with a total of four participants. The strengths reported from participants are displayed in \autoref{tab:us_good} and the problems identified during these studies are presented in \autoref{tab:us}.

\begin{table*}[h]
\centering
\scalebox{.75}{
  \begin{tabular}{l|c}
    \toprule
    \textbf{Section} & \textbf{Question} \\
    \hline
    \multirow{5}{*}{\begin{tabular}[c]{@{}c@{}} Demographic \\ Questions \end{tabular} }   &  1. How old are you?          \\
         &  2. What is the highest level of education you have completed? \\
                                      & 3. What is your native language?  \\
         &  4. Which language did you select as the theme of the game (i.e., the language whose dialectal variations you explored)?     \\
          & 5. How familiar are you with the language you explored in the game?      \\
          \hline
   \multirow{10}{*}{\begin{tabular}[c]{@{}c@{}} SUS \\ Questions \end{tabular} }     & 1. I think I would use this interface frequently.          \\
   & 2. I found the interface unnecessarily complex.         \\
    & 3. I thought the interface was easy to use. \\
    & 4. I think that I would need the support of a technical person to be able to use this interface. \\
    & 5. I found the various functions in this interface were well integrated. \\
    & 6. I thought there was too much inconsistency in this interface. \\
    & 7. I would imagine that most people would learn to use this interface very quickly. \\
    & 8. I found this interface very cumbersome to use. \\
    & 9. I felt very confident using the interface. \\
    & 10. I needed to learn a lot of things before I could get going with this interface. \\
    \hline
    \multirow{3}{*}{Feedback}      &  1. What do you like the most about Dia-Lingle?   \\        
                & 2. What do you dislike the most about Dia-Lingle?  \\
                & 3. ask for more feedback \& contributions. \\
    \bottomrule
  \end{tabular}
  }
  \caption{Detailed content of the survey distributed for usability study.}
  \label{tab:survey}
\end{table*}

\begin{table*}[h]
  \scalebox{.8}{
    \begin{tabular}{rlrrrrrr}
    \toprule
    \multicolumn{1}{c}{\textbf{Stages}} & \multicolumn{1}{c}{\textbf{Sub-stages}} & \multicolumn{1}{c}{\textbf{Duration}} & \multicolumn{5}{c}{\textbf{Description}} \\ \hline
    \multicolumn{1}{l}{Stage I} & Welcome Stage & \multicolumn{1}{c}{1 min} & \multicolumn{5}{c}{Begin by briefly introducing yourself} \\
    \multicolumn{1}{l}{Stage II} & Stage A & \multicolumn{1}{c}{2 mins} & \multicolumn{5}{c}{Semi-Structured Interview for Expectation Check} \\
          & Stage B & \multicolumn{1}{c}{2 - 3 mins} & \multicolumn{5}{c}{Explain motivations + UI design} \\
    \multicolumn{1}{l}{Stage III} & Stage C & \multicolumn{1}{c}{10 - 15 mins} & \multicolumn{5}{c}{Participants try interacting with the interface on Quiz Game and/or Match Game} \\
    \multicolumn{1}{l}{Stage IV} & Break Stage & \multicolumn{1}{c}{1 min} & \multicolumn{5}{c}{A short break} \\
    \multicolumn{1}{l}{Stage V} & Stage D & \multicolumn{1}{c}{3 - 5 mins} & \multicolumn{5}{c}{Explain models + design choices} \\
    & Stage E & \multicolumn{1}{c}{2 - 4 mins} & \multicolumn{5}{c}{Concluding Semi-Structured Interview} \\
    \multicolumn{1}{l}{Stage VI} & Stage F & \multicolumn{1}{c}{2 mins} & \multicolumn{5}{c}{Informal Questionnaire} \\
     \multicolumn{1}{l}{Stage VII} & End Stage & \multicolumn{1}{c}{1 min} & \multicolumn{5}{c}{End the study and thank the participant} \\
    \bottomrule
    \end{tabular}
    }
    \caption{Table of instructions of Usability Studies. The complete duration of the study is expected to be approximately 30 minutes. At the sub-stage F, we hand out the same survey shown in \autoref{tab:survey} to the participants.}
    \label{tab:u_s}
\end{table*}%

\begin{table*}[t]
\centering
\scalebox{.7}{
\begin{tabular}{|cllll|lllll|lllll|lllll|lllll|}
\hline
\multicolumn{5}{|l|}{\multirow{2}{*}{\textbf{Strengths reported in Usability Studies}}} & \multicolumn{5}{c|}{\textbf{P1}}  & \multicolumn{5}{c|}{\textbf{P2}}  & \multicolumn{5}{c|}{\textbf{P3}} & \multicolumn{5}{c|}{\textbf{P4}}  \\ \cline{6-25}
\multicolumn{5}{|l|}{}  & \multicolumn{1}{c|}{\textbf{E}}  & \multicolumn{1}{c|}{\textbf{C}} & \multicolumn{1}{c|}{\textbf{Q}} & \multicolumn{1}{c|}{\textbf{R}} & \multicolumn{1}{c|}{\textbf{M}} & \multicolumn{1}{c|}{\textbf{E}}  & \multicolumn{1}{c|}{\textbf{C}} & \multicolumn{1}{c|}{\textbf{Q}} & \multicolumn{1}{c|}{\textbf{R}} & \multicolumn{1}{c|}{\textbf{M}}  & \multicolumn{1}{c|}{\textbf{E}}  & \multicolumn{1}{c|}{\textbf{C}} & \multicolumn{1}{c|}{\textbf{Q}} & \multicolumn{1}{c|}{\textbf{R}} & \multicolumn{1}{c|}{\textbf{M}}  & \multicolumn{1}{c|}{\textbf{E}}  & \multicolumn{1}{c|}{\textbf{C}} & \multicolumn{1}{c|}{\textbf{Q}} & \multicolumn{1}{c|}{\textbf{R}} & \multicolumn{1}{c|}{\textbf{M}}  \\ \hline
\multicolumn{5}{|c|}{multilingual support} & \multicolumn{1}{c|}{\cellcolor[HTML]{A3CFA5}} & \multicolumn{1}{c|}{} & \multicolumn{1}{c|}{}                         & \multicolumn{1}{c|}{}                         & \multicolumn{1}{c|}{}                         & \multicolumn{1}{l|}{} & \multicolumn{1}{l|}{}   &   \multicolumn{1}{c|}{} & \multicolumn{1}{c|}{} & \multicolumn{1}{l|}{} & \multicolumn{1}{l|}{\cellcolor[HTML]{A3CFA5}}   &  \multicolumn{1}{c|}{} & \multicolumn{1}{c|}{}   & \multicolumn{1}{l|}{}                         & \multicolumn{1}{l|}{}   &      \multicolumn{1}{c|}{\cellcolor[HTML]{A3CFA5}}   & \multicolumn{1}{c|}{} & \multicolumn{1}{l|}{} & \multicolumn{1}{l|}{}  &                          \\ \hline
\multicolumn{5}{|c|}{rolling pins to catch attention}  & \multicolumn{1}{c|}{} & \multicolumn{1}{c|}{} & \multicolumn{1}{c|}{}  & \multicolumn{1}{c|}{} & \multicolumn{1}{c|}{} & \multicolumn{1}{l|}{\cellcolor[HTML]{A3CFA5}} & \multicolumn{1}{l|}{}   &  \multicolumn{1}{l|}{} & \multicolumn{1}{c|}{}                         & \multicolumn{1}{l|}{}     & \multicolumn{1}{l|}{}   &    \multicolumn{1}{l|}{}  & \multicolumn{1}{c|}{} & \multicolumn{1}{l|}{}                         & \multicolumn{1}{l|}{}   &      \multicolumn{1}{l|}{\cellcolor[HTML]{A3CFA5}}           & \multicolumn{1}{c|}{} & \multicolumn{1}{l|}{} & \multicolumn{1}{l|}{}                         &                          \\ \hline
\multicolumn{5}{|c|}{interface is very transparent}   & \multicolumn{1}{l|}{}      & \multicolumn{1}{l|}{\cellcolor[HTML]{A3CFA5}}                         & \multicolumn{1}{l|}{}                         &  \multicolumn{1}{l|}{}                     & \multicolumn{1}{l|}{}                         & \multicolumn{1}{l|}{} & \multicolumn{1}{l|}{}   &     \multicolumn{1}{l|}{\cellcolor[HTML]{A3CFA5}}                      & \multicolumn{1}{l|}{}                         & \multicolumn{1}{l|}{\cellcolor[HTML]{A3CFA5}}                         & \multicolumn{1}{l|}{}   &             \multicolumn{1}{l|}{}              & \multicolumn{1}{l|}{}                         & \multicolumn{1}{l|}{}                         & \multicolumn{1}{l|}{\cellcolor[HTML]{A3CFA5}}   &           \multicolumn{1}{l|}{}               & \multicolumn{1}{l|}{}    & \multicolumn{1}{l|}{} & \multicolumn{1}{l|}{}  &  \\ \hline
\multicolumn{5}{|c|}{animation of component transition}   & \multicolumn{1}{l|}{}      & \multicolumn{1}{l|}{}                         & \multicolumn{1}{l|}{\cellcolor[HTML]{A3CFA5}}                         &  \multicolumn{1}{l|}{}                     & \multicolumn{1}{l|}{}                         & \multicolumn{1}{l|}{} & \multicolumn{1}{l|}{}   &     \multicolumn{1}{l|}{\cellcolor[HTML]{A3CFA5}}                      & \multicolumn{1}{l|}{}                         & \multicolumn{1}{l|}{}                         & \multicolumn{1}{l|}{}   &             \multicolumn{1}{l|}{}              & \multicolumn{1}{l|}{}                         & \multicolumn{1}{l|}{}                         & \multicolumn{1}{l|}{}   &           \multicolumn{1}{l|}{}               & \multicolumn{1}{l|}{}                         & \multicolumn{1}{l|}{\cellcolor[HTML]{A3CFA5}} & \multicolumn{1}{l|}{}  &  \\ \hline
\multicolumn{5}{|c|}{hovering effects}   & \multicolumn{1}{l|}{}      & \multicolumn{1}{l|}{}                         & \multicolumn{1}{l|}{\cellcolor[HTML]{A3CFA5}}                         &  \multicolumn{1}{l|}{}                     & \multicolumn{1}{l|}{\cellcolor[HTML]{A3CFA5}}                         & \multicolumn{1}{l|}{\cellcolor[HTML]{A3CFA5}} & \multicolumn{1}{l|}{}   &     \multicolumn{1}{l|}{}                      & \multicolumn{1}{l|}{}                         & \multicolumn{1}{l|}{}                         & \multicolumn{1}{l|}{\cellcolor[HTML]{A3CFA5}}   &             \multicolumn{1}{l|}{}              & \multicolumn{1}{l|}{}                         & \multicolumn{1}{l|}{}                         & \multicolumn{1}{l|}{}   &           \multicolumn{1}{l|}{\cellcolor[HTML]{A3CFA5}}               & \multicolumn{1}{l|}{}                         & \multicolumn{1}{l|}{\cellcolor[HTML]{A3CFA5}} & \multicolumn{1}{l|}{}  &  \\ \hline
\multicolumn{5}{|c|}{\begin{tabular}[c]{@{}c@{}}pop-up window to prevent \\ from accidental click \end{tabular}} & \multicolumn{1}{l|}{} & \multicolumn{1}{l|}{}                         & \multicolumn{1}{l|}{} &   \multicolumn{1}{l|}{}      & \multicolumn{1}{l|}{}       & \multicolumn{1}{l|}{}                         & \multicolumn{1}{l|}{}   &       \multicolumn{1}{l|}{\cellcolor[HTML]{A3CFA5}}    & \multicolumn{1}{l|}{}      & \multicolumn{1}{l|}{}    & \multicolumn{1}{l|}{}   & \multicolumn{1}{l|}{}  & \multicolumn{1}{l|}{\cellcolor[HTML]{A3CFA5}}                         & \multicolumn{1}{l|}{}                         & \multicolumn{1}{l|}{}   &   \multicolumn{1}{l|}{}   & \multicolumn{1}{l|}{} & \multicolumn{1}{l|}{}                         & \multicolumn{1}{l|}{\cellcolor[HTML]{A3CFA5}} & \\ \hline
\multicolumn{5}{|c|}{smooth transition of components}   & \multicolumn{1}{l|}{}      & \multicolumn{1}{l|}{}                         & \multicolumn{1}{l|}{}                         &  \multicolumn{1}{l|}{\cellcolor[HTML]{A3CFA5}}                     & \multicolumn{1}{l|}{}                         & \multicolumn{1}{l|}{} & \multicolumn{1}{l|}{}   &     \multicolumn{1}{l|}{}                      & \multicolumn{1}{l|}{}                         & \multicolumn{1}{l|}{}                         & \multicolumn{1}{l|}{}   &             \multicolumn{1}{l|}{}              & \multicolumn{1}{l|}{}                         & \multicolumn{1}{l|}{\cellcolor[HTML]{A3CFA5}}                         & \multicolumn{1}{l|}{}   &           \multicolumn{1}{l|}{}               & \multicolumn{1}{l|}{}                         & \multicolumn{1}{l|}{} & \multicolumn{1}{l|}{}  &  \\ \hline
\multicolumn{5}{|c|}{typing suggestions}   & \multicolumn{1}{l|}{}      & \multicolumn{1}{l|}{}                         & \multicolumn{1}{l|}{}                         &  \multicolumn{1}{l|}{}                     & \multicolumn{1}{l|}{}                         & \multicolumn{1}{l|}{} & \multicolumn{1}{l|}{}   &     \multicolumn{1}{l|}{\cellcolor[HTML]{A3CFA5}}                      & \multicolumn{1}{l|}{}                         & \multicolumn{1}{l|}{}                         & \multicolumn{1}{l|}{}   &             \multicolumn{1}{l|}{}              & \multicolumn{1}{l|}{}                         & \multicolumn{1}{l|}{}                         & \multicolumn{1}{l|}{}   &           \multicolumn{1}{l|}{}               & \multicolumn{1}{l|}{}                         & \multicolumn{1}{l|}{\cellcolor[HTML]{A3CFA5}} & \multicolumn{1}{l|}{}  &  \\ \hline
\multicolumn{5}{|c|}{dialect visualisation on map}   & \multicolumn{1}{l|}{}      & \multicolumn{1}{l|}{}                         & \multicolumn{1}{l|}{}                         &  \multicolumn{1}{l|}{}                     & \multicolumn{1}{l|}{}                         & \multicolumn{1}{l|}{} & \multicolumn{1}{l|}{}   &     \multicolumn{1}{l|}{}                      & \multicolumn{1}{l|}{\cellcolor[HTML]{A3CFA5}}                         & \multicolumn{1}{l|}{}                         & \multicolumn{1}{l|}{}   &             \multicolumn{1}{l|}{}              & \multicolumn{1}{l|}{}                         & \multicolumn{1}{l|}{\cellcolor[HTML]{A3CFA5}}                         & \multicolumn{1}{l|}{}   &           \multicolumn{1}{l|}{}               & \multicolumn{1}{l|}{}                         & \multicolumn{1}{l|}{} & \multicolumn{1}{l|}{}  &  \\ \hline
\multicolumn{5}{|c|}{gamified difficulty levels}   & \multicolumn{1}{l|}{}      & \multicolumn{1}{l|}{}                         & \multicolumn{1}{l|}{\cellcolor[HTML]{A3CFA5}}                         &  \multicolumn{1}{l|}{}                     & \multicolumn{1}{l|}{}                         & \multicolumn{1}{l|}{} & \multicolumn{1}{l|}{}   &     \multicolumn{1}{l|}{\cellcolor[HTML]{A3CFA5}}                      & \multicolumn{1}{l|}{}                         & \multicolumn{1}{l|}{}                         & \multicolumn{1}{l|}{}   &             \multicolumn{1}{l|}{}              & \multicolumn{1}{l|}{\cellcolor[HTML]{A3CFA5}}                         & \multicolumn{1}{l|}{}                         & \multicolumn{1}{l|}{}   &           \multicolumn{1}{l|}{}               & \multicolumn{1}{l|}{}                         & \multicolumn{1}{l|}{} & \multicolumn{1}{l|}{}  &  \\ \hline
\multicolumn{5}{|c|}{onboarding guidelines for lasso tool}   & \multicolumn{1}{l|}{}      & \multicolumn{1}{l|}{}                         & \multicolumn{1}{l|}{}                         &  \multicolumn{1}{l|}{}                     & \multicolumn{1}{l|}{}                         & \multicolumn{1}{l|}{} & \multicolumn{1}{l|}{}   &     \multicolumn{1}{l|}{}                      & \multicolumn{1}{l|}{}                         & \multicolumn{1}{l|}{}                         & \multicolumn{1}{l|}{}   &             \multicolumn{1}{l|}{}              & \multicolumn{1}{l|}{}                         & \multicolumn{1}{l|}{\cellcolor[HTML]{A3CFA5}}                         & \multicolumn{1}{l|}{}   &           \multicolumn{1}{l|}{}               & \multicolumn{1}{l|}{}                         & \multicolumn{1}{l|}{} & \multicolumn{1}{l|}{\cellcolor[HTML]{A3CFA5}}  &  \\ \hline
\multicolumn{5}{|c|}{classifier is relatively robust}   & \multicolumn{1}{l|}{}      & \multicolumn{1}{l|}{}                         & \multicolumn{1}{l|}{}                         &  \multicolumn{1}{l|}{}                     & \multicolumn{1}{l|}{}                         & \multicolumn{1}{l|}{} & \multicolumn{1}{l|}{}   &     \multicolumn{1}{l|}{}                      & \multicolumn{1}{l|}{\cellcolor[HTML]{A3CFA5}}                         & \multicolumn{1}{l|}{}                         & \multicolumn{1}{l|}{}   &             \multicolumn{1}{l|}{}              & \multicolumn{1}{l|}{}                         & \multicolumn{1}{l|}{\cellcolor[HTML]{A3CFA5}}                         & \multicolumn{1}{l|}{}   &           \multicolumn{1}{l|}{}               & \multicolumn{1}{l|}{}                         & \multicolumn{1}{l|}{} & \multicolumn{1}{l|}{\cellcolor[HTML]{A3CFA5}}  & \multicolumn{1}{l|}{} \\ \hline
\multicolumn{5}{|c|}{the overall idea}   & \multicolumn{1}{l|}{}      & \multicolumn{1}{l|}{}                         & \multicolumn{1}{l|}{}                         &  \multicolumn{1}{l|}{}                     & \multicolumn{1}{l|}{}                         & \multicolumn{1}{l|}{\cellcolor[HTML]{A3CFA5}} & \multicolumn{1}{l|}{\cellcolor[HTML]{A3CFA5}}   &     \multicolumn{1}{l|}{\cellcolor[HTML]{A3CFA5}}                      & \multicolumn{1}{l|}{\cellcolor[HTML]{A3CFA5}}                         & \multicolumn{1}{l|}{\cellcolor[HTML]{A3CFA5}}                         & \multicolumn{1}{l|}{\cellcolor[HTML]{A3CFA5}}   &             \multicolumn{1}{l|}{\cellcolor[HTML]{A3CFA5}}              & \multicolumn{1}{l|}{\cellcolor[HTML]{A3CFA5}}                         & \multicolumn{1}{l|}{\cellcolor[HTML]{A3CFA5}}                         & \multicolumn{1}{l|}{\cellcolor[HTML]{A3CFA5}}   &           \multicolumn{1}{l|}{\cellcolor[HTML]{A3CFA5}}               & \multicolumn{1}{l|}{\cellcolor[HTML]{A3CFA5}}                         & \multicolumn{1}{l|}{\cellcolor[HTML]{A3CFA5}} & \multicolumn{1}{l|}{\cellcolor[HTML]{A3CFA5}}  & \multicolumn{1}{l|}{\cellcolor[HTML]{A3CFA5}} \\ \hline

\end{tabular}
}
  \caption{A detailed distribution matrix of usability strengths showing which spotlights found in which step by which participant in Usability Studies. \textbf{P1}; \textbf{P2}; \textbf{P3}; \textbf{P4} represent four participants. \textbf{E}; \textbf{C}; \textbf{Q}; \textbf{R} and \textbf{M} represent different stages of \textit{Entry}; \textit{Choice}; \textit{Quiz}; \textit{Review} and \textit{Match} on dashboard detailed in Section~\ref{sec:interface_design}.}
  \label{tab:us_good}
\end{table*}

\begin{table*}[t]
\centering
\scalebox{.7}{
\begin{tabular}{|cllll|lllll|lllll|lllll|lllll|}
\hline
\multicolumn{5}{|l|}{\multirow{2}{*}{\textbf{Problems reported in Usability Studies}}} & \multicolumn{5}{c|}{\textbf{P1}}  & \multicolumn{5}{c|}{\textbf{P2}}  & \multicolumn{5}{c|}{\textbf{P3}} & \multicolumn{5}{c|}{\textbf{P4}}  \\ \cline{6-25}
\multicolumn{5}{|l|}{}  & \multicolumn{1}{c|}{\textbf{E}}  & \multicolumn{1}{c|}{\textbf{C}} & \multicolumn{1}{c|}{\textbf{Q}} & \multicolumn{1}{c|}{\textbf{R}} & \multicolumn{1}{c|}{\textbf{M}} & \multicolumn{1}{c|}{\textbf{E}}  & \multicolumn{1}{c|}{\textbf{C}} & \multicolumn{1}{c|}{\textbf{Q}} & \multicolumn{1}{c|}{\textbf{R}} & \multicolumn{1}{c|}{\textbf{M}}  & \multicolumn{1}{c|}{\textbf{E}}  & \multicolumn{1}{c|}{\textbf{C}} & \multicolumn{1}{c|}{\textbf{Q}} & \multicolumn{1}{c|}{\textbf{R}} & \multicolumn{1}{c|}{\textbf{M}}  & \multicolumn{1}{c|}{\textbf{E}}  & \multicolumn{1}{c|}{\textbf{C}} & \multicolumn{1}{c|}{\textbf{Q}} & \multicolumn{1}{c|}{\textbf{R}} & \multicolumn{1}{c|}{\textbf{M}}  \\ \hline
\multicolumn{5}{|c|}{some components are small} & \multicolumn{1}{c|}{\cellcolor[HTML]{E1BC81}} & \multicolumn{1}{c|}{} & \multicolumn{1}{c|}{}                         & \multicolumn{1}{c|}{}                         & \multicolumn{1}{c|}{}                         & \multicolumn{1}{l|}{} & \multicolumn{1}{l|}{}   &   \multicolumn{1}{c|}{} & \multicolumn{1}{c|}{} & \multicolumn{1}{l|}{} & \multicolumn{1}{l|}{}   &  \multicolumn{1}{c|}{} & \multicolumn{1}{c|}{}   & \multicolumn{1}{l|}{}                         & \multicolumn{1}{l|}{}   &      \multicolumn{1}{c|}{\cellcolor[HTML]{E1BC81}}   & \multicolumn{1}{c|}{} & \multicolumn{1}{l|}{} & \multicolumn{1}{l|}{}  &                          \\ \hline
\multicolumn{5}{|c|}{unclear about how to proceed}  & \multicolumn{1}{c|}{} & \multicolumn{1}{c|}{\cellcolor[HTML]{E1BC81}} & \multicolumn{1}{c|}{\cellcolor[HTML]{E1BC81}}  & \multicolumn{1}{c|}{} & \multicolumn{1}{c|}{} & \multicolumn{1}{l|}{} & \multicolumn{1}{l|}{}   &  \multicolumn{1}{l|}{} & \multicolumn{1}{c|}{}                         & \multicolumn{1}{l|}{}     & \multicolumn{1}{l|}{}   &    \multicolumn{1}{l|}{}  & \multicolumn{1}{c|}{\cellcolor[HTML]{E1BC81}} & \multicolumn{1}{l|}{}                         & \multicolumn{1}{l|}{}   &      \multicolumn{1}{l|}{}           & \multicolumn{1}{c|}{} & \multicolumn{1}{l|}{} & \multicolumn{1}{l|}{}                         &                          \\ \hline
\multicolumn{5}{|c|}{feel \textit{Match} is harder than \textit{Quiz}}   & \multicolumn{1}{l|}{}      & \multicolumn{1}{l|}{}                         & \multicolumn{1}{l|}{}                         &  \multicolumn{1}{l|}{}                     & \multicolumn{1}{l|}{}                         & \multicolumn{1}{l|}{} & \multicolumn{1}{l|}{}   &     \multicolumn{1}{l|}{\cellcolor[HTML]{E1BC81}}                      & \multicolumn{1}{l|}{}                         & \multicolumn{1}{l|}{\cellcolor[HTML]{E1BC81}}                         & \multicolumn{1}{l|}{}   &             \multicolumn{1}{l|}{}              & \multicolumn{1}{l|}{}                         & \multicolumn{1}{l|}{}                         & \multicolumn{1}{l|}{}   &           \multicolumn{1}{l|}{}               & \multicolumn{1}{l|}{}                         & \multicolumn{1}{l|}{} & \multicolumn{1}{l|}{}  &  \\ \hline
\multicolumn{5}{|c|}{\begin{tabular}[c]{@{}c@{}}click the components before \\ onboarding guidelines finish\end{tabular}} & \multicolumn{1}{l|}{} & \multicolumn{1}{l|}{}                         & \multicolumn{1}{l|}{} &   \multicolumn{1}{l|}{}      & \multicolumn{1}{l|}{}       & \multicolumn{1}{l|}{}                         & \multicolumn{1}{l|}{}   &       \multicolumn{1}{l|}{}    & \multicolumn{1}{l|}{}      & \multicolumn{1}{l|}{}    & \multicolumn{1}{l|}{}   & \multicolumn{1}{l|}{}  & \multicolumn{1}{l|}{}                         & \multicolumn{1}{l|}{}                         & \multicolumn{1}{l|}{}   &   \multicolumn{1}{l|}{}   & \multicolumn{1}{l|}{} & \multicolumn{1}{l|}{}                         & \multicolumn{1}{l|}{\cellcolor[HTML]{E1BC81}} & \\ \hline
\multicolumn{5}{|c|}{do not like the text input blocks}   & \multicolumn{1}{l|}{}      & \multicolumn{1}{l|}{}                         & \multicolumn{1}{l|}{\cellcolor[HTML]{E1BC81}}                         &  \multicolumn{1}{l|}{}                     & \multicolumn{1}{l|}{}                         & \multicolumn{1}{l|}{} & \multicolumn{1}{l|}{}   &     \multicolumn{1}{l|}{}                      & \multicolumn{1}{l|}{}                         & \multicolumn{1}{l|}{}                         & \multicolumn{1}{l|}{}   &             \multicolumn{1}{l|}{}              & \multicolumn{1}{l|}{\cellcolor[HTML]{E1BC81}}                         & \multicolumn{1}{l|}{}                         & \multicolumn{1}{l|}{}   &           \multicolumn{1}{l|}{}               & \multicolumn{1}{l|}{}                         & \multicolumn{1}{l|}{} & \multicolumn{1}{l|}{}  &  \\ \hline
\multicolumn{5}{|c|}{do not understand colour legend}   & \multicolumn{1}{l|}{}      & \multicolumn{1}{l|}{}                         & \multicolumn{1}{l|}{}                         &  \multicolumn{1}{l|}{}                     & \multicolumn{1}{l|}{}                         & \multicolumn{1}{l|}{} & \multicolumn{1}{l|}{}   &     \multicolumn{1}{l|}{}                      & \multicolumn{1}{l|}{}                         & \multicolumn{1}{l|}{}                         & \multicolumn{1}{l|}{}   &             \multicolumn{1}{l|}{}              & \multicolumn{1}{l|}{}                         & \multicolumn{1}{l|}{}                         & \multicolumn{1}{l|}{}   &           \multicolumn{1}{l|}{}               & \multicolumn{1}{l|}{}                         & \multicolumn{1}{l|}{\cellcolor[HTML]{E1BC81}} & \multicolumn{1}{l|}{}  &  \\ \hline
\multicolumn{5}{|c|}{do not understand probability}   & \multicolumn{1}{l|}{}      & \multicolumn{1}{l|}{}                         & \multicolumn{1}{l|}{}                         &  \multicolumn{1}{l|}{\cellcolor[HTML]{E1BC81}}                     & \multicolumn{1}{l|}{}                         & \multicolumn{1}{l|}{} & \multicolumn{1}{l|}{}   &     \multicolumn{1}{l|}{}                      & \multicolumn{1}{l|}{}                         & \multicolumn{1}{l|}{}                         & \multicolumn{1}{l|}{}   &             \multicolumn{1}{l|}{}              & \multicolumn{1}{l|}{}                         & \multicolumn{1}{l|}{}                         & \multicolumn{1}{l|}{}   &           \multicolumn{1}{l|}{}               & \multicolumn{1}{l|}{}                         & \multicolumn{1}{l|}{} & \multicolumn{1}{l|}{}  &  \\ \hline
\multicolumn{5}{|c|}{want more explanations on prediction}   & \multicolumn{1}{l|}{}      & \multicolumn{1}{l|}{}                         & \multicolumn{1}{l|}{}                         &  \multicolumn{1}{l|}{}                     & \multicolumn{1}{l|}{}                         & \multicolumn{1}{l|}{} & \multicolumn{1}{l|}{}   &     \multicolumn{1}{l|}{}                      & \multicolumn{1}{l|}{\cellcolor[HTML]{E1BC81}}                         & \multicolumn{1}{l|}{}                         & \multicolumn{1}{l|}{}   &             \multicolumn{1}{l|}{}              & \multicolumn{1}{l|}{}                         & \multicolumn{1}{l|}{}                         & \multicolumn{1}{l|}{}   &           \multicolumn{1}{l|}{}               & \multicolumn{1}{l|}{}                         & \multicolumn{1}{l|}{} & \multicolumn{1}{l|}{}  &  \\ \hline
\multicolumn{5}{|c|}{yellow ticks are confusing}   & \multicolumn{1}{l|}{}      & \multicolumn{1}{l|}{}                         & \multicolumn{1}{l|}{}                         &  \multicolumn{1}{l|}{}                     & \multicolumn{1}{l|}{\cellcolor[HTML]{E1BC81}}                         & \multicolumn{1}{l|}{} & \multicolumn{1}{l|}{}   &     \multicolumn{1}{l|}{}                      & \multicolumn{1}{l|}{}                         & \multicolumn{1}{l|}{\cellcolor[HTML]{E1BC81}}                         & \multicolumn{1}{l|}{}   &             \multicolumn{1}{l|}{}              & \multicolumn{1}{l|}{}                         & \multicolumn{1}{l|}{}                         & \multicolumn{1}{l|}{}   &           \multicolumn{1}{l|}{}               & \multicolumn{1}{l|}{}                         & \multicolumn{1}{l|}{} & \multicolumn{1}{l|}{}  &  \\ \hline
\multicolumn{5}{|c|}{do not understand the standardised sentence}   & \multicolumn{1}{l|}{}      & \multicolumn{1}{l|}{}                         & \multicolumn{1}{l|}{\cellcolor[HTML]{E1BC81}}                         &  \multicolumn{1}{l|}{}                     & \multicolumn{1}{l|}{}                         & \multicolumn{1}{l|}{} & \multicolumn{1}{l|}{}   &     \multicolumn{1}{l|}{}                      & \multicolumn{1}{l|}{}                         & \multicolumn{1}{l|}{}                         & \multicolumn{1}{l|}{}   &             \multicolumn{1}{l|}{}              & \multicolumn{1}{l|}{}                         & \multicolumn{1}{l|}{}                         & \multicolumn{1}{l|}{}   &           \multicolumn{1}{l|}{}               & \multicolumn{1}{l|}{}                         & \multicolumn{1}{l|}{} & \multicolumn{1}{l|}{}  &  \\ \hline
\multicolumn{5}{|c|}{want audio input option}   & \multicolumn{1}{l|}{}      & \multicolumn{1}{l|}{}                         & \multicolumn{1}{l|}{}                         &  \multicolumn{1}{l|}{}                     & \multicolumn{1}{l|}{}                         & \multicolumn{1}{l|}{} & \multicolumn{1}{l|}{}   &     \multicolumn{1}{l|}{\cellcolor[HTML]{E1BC81}}                      & \multicolumn{1}{l|}{}                         & \multicolumn{1}{l|}{}                         & \multicolumn{1}{l|}{}   &             \multicolumn{1}{l|}{}              & \multicolumn{1}{l|}{}                         & \multicolumn{1}{l|}{}                         & \multicolumn{1}{l|}{}   &           \multicolumn{1}{l|}{}               & \multicolumn{1}{l|}{}                         & \multicolumn{1}{l|}{\cellcolor[HTML]{E1BC81}} & \multicolumn{1}{l|}{}  & \multicolumn{1}{l|}{\cellcolor[HTML]{E1BC81}} \\ \hline
\multicolumn{5}{|c|}{want instruction texts larger}   & \multicolumn{1}{l|}{}      & \multicolumn{1}{l|}{}                         & \multicolumn{1}{l|}{\cellcolor[HTML]{E1BC81}}                         &  \multicolumn{1}{l|}{}                     & \multicolumn{1}{l|}{}                         & \multicolumn{1}{l|}{} & \multicolumn{1}{l|}{}   &     \multicolumn{1}{l|}{}                      & \multicolumn{1}{l|}{}                         & \multicolumn{1}{l|}{}                         & \multicolumn{1}{l|}{}   &             \multicolumn{1}{l|}{}              & \multicolumn{1}{l|}{}                         & \multicolumn{1}{l|}{}                         & \multicolumn{1}{l|}{}   &           \multicolumn{1}{l|}{}               & \multicolumn{1}{l|}{}                         & \multicolumn{1}{l|}{} & \multicolumn{1}{l|}{}  & \multicolumn{1}{l|}{} \\ \hline
\end{tabular}
}
  \caption{A detailed distribution matrix of usability problems showing which problems found in which step by which participant in Usability Studies. \textbf{P1}; \textbf{P2}; \textbf{P3}; \textbf{P4} represent four participants. \textbf{E}; \textbf{C}; \textbf{Q}; \textbf{R} and \textbf{M} represent different stages of \textit{Entry}; \textit{Choice}; \textit{Quiz}; \textit{Review} and \textit{Match} on dashboard detailed in Section~\ref{sec:interface_design}.}
  \label{tab:us}
  \vspace{5cm}
\end{table*}

\end{document}